\documentstyle[pubform]{article}

\renewcommand{\baselinestretch}{.75}

\title{\LARGE\bf On Assessing the Complexity of Software Architectures}
\author{{\bf Jianjun Zhao}\\
\large Department of Computer Science and Engineering\\
\large Fukuoka Institute of Technology\\
\large 3-10-1 Wajiro-Higashi, Higashi-ku, Fukuoka 811-0214, Japan\\
\large zhao@cs.fit.ac.jp}

\date{}

\begin{document}
\maketitle

\abstract{
This paper proposes some new architectural metrics which are 
appropriate for evaluating the architectural attributes of a software 
system. The main feature of our approach is to assess the 
complexity of a software architecture by analyzing 
various types of architectural dependences in the architecture.}

\medskip
\noindent
{\bf keywords:} Architectural description lanaguage, Architectural metric, 
Dependence analysis, Software architecture

\section{Introduction}
\label{sec:intro}

Software metrics have many applications in software engineering 
activities including software analysis, testing, debugging,
maintenance, and project management.
In the past two decades numerous software metrics 
have been proposed for measuring the complexity of software
\cite{Fenton97,Zuse90}. These metrics can be divided into two 
categories according to the design levels of software\footnote{There are usually two levels of design for software, {\it architectural level design} 
where involves overall association of system capability with components, 
and {\it code level design} where involves algorithms 
and data structures \cite{Rombach90}.}: {\it code metrics} which aim at
measuring the complexity of a single program module at
code design level \cite{Cheng93,Halstead77,Harrison87,Mccabe76}, 
and {\it architectural metrics} 
which aim at measuring the complexity of components and their
interconnections in software systems at architectural design level
\cite{Henry81,Kazman98,Mccabe89,Yin78}. 

Most work on software metrics focused on code metrics
which are derived solely from source code of a program, and 
the study of architectural metrics has received little attention. 
However, architectural measurement can be regarded as a desirable 
addition to code metrics because it allows you to capture important 
aspects of a system's architecture earlier in the system life cycle so
you can take corrective actions earlier \cite{Rombach90}. This may 
offer greater potential for return on investment in order to make
large gains in productivity and quality since error detection and
repair is more costly if we can not catch errors in the early stage 
of system design.

But, why the study of architectural metrics has received little
attention in comparison with code metrics ? 
One important reason is while the code level for software
systems is now well understood, the architectural level is currently 
understood mostly at the level of intuition, anecdote, and 
folklore \cite{Shaw96}. 
Existing representations that a system architect uses to represent the 
architecture of a software system are usually informal and {\it ad hoc}, 
and therefore can not capture enough useful information of the
system's architecture. Moreover, with such an informal and {\it ad hoc} manner, it is
difficult to develop analysis tools to automatically support the
evaluation and comparison of existing architectural metrics. As a
result, in order to make
architectural metrics more widely accepted and used in software
system design, formal representation of system architectures is
strongly needed. 

Recently, as the size and complexity of software systems increases, the design
and specification of the overall software architecture of a system is 
receiving increasingly attention. The software architecture of a
system defines its high-level structure, exposing
its gross organization as a collection of interacting components.
A well-defined architecture allows an engineer to reason about system
properties at a high level of abstraction \cite{Shaw96}.
Architecture description languages (ADLs) are formal languages that
can be used to represent the architecture of a software system.
They focus on the high-level structure of the overall application
rather than the implementation details of any specific source module.
In order to support formal representation and reasoning of software 
architecture, a number of ADLs such as W{\sc right} \cite{Allen97}, 
Rapide \cite {Luckham95}, and UniCon
\cite{Shaw95} have been proposed. By using an ADL, a system architect 
can formally represent various general attributes of a software system's 
architecture. This provides researchers with a promising solution to 
solve the problems existing in recent architectural metrics. First, 
a sound basis for software architecture promises one to define new 
architectural metrics, or refine existing architectural metrics in a more 
formal way in comparing with existing informal structure charts based 
architectural metrics. Second, formal language support for software 
architecture provides a useful platform on which automated support tools
for architectural metrics can be developed and formal 
evaluation and comparison of existing architectural metrics can be 
done.

In this paper, we propose some new architectural metrics for
software architecture. Our metrics are appropriate for 
evaluating the architectural attributes of a software system. The main 
feature of our approach is to assess the complexity of a
software architecture by analyzing various types of architectural dependences
in a software architecture. To formally define these metrics, we present a 
dependence-based representation named {\em Architectural Dependence 
Graph} (ADG) to explicitly represent various architectural dependences in 
a software architecture. 

The rest of the paper is organized as follows. 
Section \ref{sec:depen} presents three types of architectural
dependences in a software 
architecture and the architectural dependence graph. Section 
\ref{sec:measure} defines some dependence-based metrics for software
architecture. Section \ref{sec:work} discusses some related work. 
Concluding remarks are given in Section \ref{sec:final}.

\section{A Dependence Model for Software Architecture}
\label{sec:depen}

When we intend to measure some attributes of an entity, we must build 
some model for the entity such that the attributes can
be explicitly described in the model. In this section, we present 
a dependence model for software architecture to capture attributes 
concerning about information flow in a software architecture.

\subsection{Program Dependences}

Program dependences are dependence relationships holding between
program statements (variables) in a program that are implicitly
determined by control flow and data flow in the program. Usually, 
there are two types of program dependences in a program, that is, 
{\it control dependences} representing the control conditions on which
the execution of a statement or expression depends and {\it data
dependences} representing the flow of data between statements or 
expressions. The task to determine a program's dependences is 
called {\it program dependence analysis}. 

Program dependence analysis has been primarily studied
in the context of conventional programming languages.
In such languages, it is typically performed
using a {\it program dependence graph} \cite{Cheng93,Horwitz90,Ottenstein84}. 
Program dependence analysis, though originally proposed for complier
optimization, has also many applications in software engineering
activities such as program slicing, understanding, debugging, testing,
maintenance and complexity measurement 
\cite{Horwitz90,Ottenstein84,Podgurski90}. As a result, 
it seems reasonable to apply program dependence analysis technique 
to software architectures to support software architecture development
activities \cite{Stafford98,Zhao97}.


\subsection{Architectural Dependences}

Roughly speaking, architectural dependences are dependence
relationships holding between components (ports) in a software architecture, 
and are implicitly determined by information flow in the architecture.
Unlike program dependences, which are defined as dependence
relationships between statements (variables) in a program, 
architectural dependences are defined as
dependence relationships between components (ports) in a software
architecture. To perform dependence analysis on software
architectures, it is important to identify all primary architectural 
dependence relationships between components (ports) in the architectures. 
However, such a work is quite difficult because comparing with 
program dependence analysis, the dependence relationships between 
components (ports) in a software architecture can be more complex
and broad. In this section we introduce three types of primary
architectural dependences between components (ports) in a software 
architecture. The classification of architectural dependence types 
based on the results of {\it coordination theory} \cite{Malone94} 
\footnote{In \cite{Malone94}, 
Malone and Crowston defines {\it coordination} as the process of 
{\it managing dependences} among activities.}. 
The types of primary architectural dependences are not limited to
these three ones, rather, new types of primary architectural
dependences must be further exploited in order to identify all types 
of primary architectural dependences in a software architecture.

\vspace{2mm}
\noindent
{\it Shared Dependences}

\vspace{2mm}
\noindent
Sharing dependences represent dependence relationships \\among 
consumers who use the same resource or producers who produce for
the same consumers. For example, for two components $u$ and $v$,
if $u$ and $v$ refer to the same global data, then there exists a
shared dependence relationship between $u$ and $v$.

\vspace{2mm}
\noindent
{\it Flow Dependences}

\vspace{2mm}
\noindent
Flow dependences represent dependence relationships
between producers and consumers of resources. For example, for two
components $u$ and $v$, if $u$ must complete before control flows into
$v$ (prerequisite), or if $u$ communicate $v$ by parameters, 
then there exists a flow dependence relationship 
between $u$ and $v$.

\vspace{2mm}
\noindent
{\it Constrained Dependences}

\vspace{2mm}
\noindent
Constrained dependences represent constraints on the relative
flow of control among a set of activities. For example, for two
components $u$ and $v$, $u$ and $v$ can not execute at the same time
(mutual exclusion), then there exists a constrained dependence
relationship between $u$ and $v$.

\subsection{Architectural Dependence Graph}
\label{sec:adg}

We present an arc-classified digraph named
{\it Architectural Dependence Graph} (ADG) for explicitly representing the
three types of primary architectural dependences in a software
architecture. Here we assume that the interface of
each component in a software architecture is defined by a set of
{\it ports}. The ADG of a software architecture consists of vertices 
and arcs to connect these vertices.
There is a {\it component vertex} for each component in the
architecture, and each component vertex consists of a set of 
{\it port vertices} each representing a port of the component. 
There is an architectural dependence arc between two port vertices 
of components if there exists a shared, flow, or constrained
dependence relationship between the ports. 

Architectural dependence information can be inferred 
based on formal architectural specifications of a software
architecture. For example, based on a W{\sc right} architectural
specification we can infer which ports of a component are input ports 
and which are output ports in the specification. 
Moreover, the direction in which the information transfers between 
ports can also be inferred based on the formal specification. Such 
kinds of information can be used to construct the architectural
dependence graph for a software architecture to formally define 
dependence-based architectural metrics.

\section{Architectural Metrics}
\label{sec:measure}

As we mentioned in Section \ref{sec:depen}, 
architectural dependences are dependence relationships holding between
components in a software architecture that are implicitly determined 
by information flow in the architecture. Therefore, architectural 
dependences can be regarded as one of intrinsic attributes of a
software architecture and it is reasonable to regard architectural
dependences as one of objects for measuring the architectural
complexity of a software architecture.

In this section, we define a set of new architectural metrics 
in terms of architectural dependences to assess the complexity 
of a software architecture from various different viewpoints. 
Once the ADG of a software architecture is constructed, 
the metrics can be computed easily based on the graph. The following 
notations are used for defining these metrics:

\begin{itemize}
	\item[] $|A|$:  the cardinality of set $A$.
	\item[] $R^+$:  the transitive closure of binary relation $R$.
	\item[] ${\sigma}_{[1]=v}(R)$: the selection of binary relation
$R$ such that ${\sigma}_{[1]=v}(R)=\{(v1,v2)| (v1,v2)\in R\ and\ v1=v\}$.    
\end{itemize}

When we constructed the ADG for a software architecture, the most
general metric can be defined in terms
of ADG. The following metric is defined for measuring the total
complexity of a software architecture: 

\begin{itemize}
\item Let $D_{t}$ be the set of all dependences arcs in the ADG of a
software architecture, then the total complexity $M_{T}$ of the 
architecture can be measured by $M_{T}= |D_{t}|.$
\end{itemize}

Note that the above metric was defined under the situation that we
treat a component as an unit to construct the ADG of a software
 architecture. However, in fact, each component in the architecture 
may generally correspond to a single application module which can be 
measured by usual code metrics at code level. So there is a need to 
combine the total complexity at architectural design level with
internal component complexity at code design level to obtain an
overall complexity metric.

\begin{itemize} 
\item Let $M_{T}$ be the total complexity and $M_{1}, ..., M_{k}$ be
the individual component complexities. Then the global complexity
$M_{G}$ of a software architecture can be measured by: 
$M_{G}= M_{T} + \sum_{i=1}^{k}M_{i}$.
\end{itemize}

The above metrics only concerned with the direct architectural
dependences in a software architecture, but did not take indirect 
architectural dependences into account. 
As a result, they only capture the sum of some local complexity, rather
than the total complexity of the architecture. In fact, a component 
in a software architecture may indirectly depend on other components
in the architecture. Therefore, to assess the total complexity of a 
software architecture, we should define a metric by taking either direct or
indirect architectural dependences into account. This can be obtained
by calculating the transitive closure $|D_{t}^+|$ of the $|D_{t}|$, 
we have: $M'_{T}= |D_{t}^+|.$ Similarly, if we also consider the
indirect dependences at
architectural level and each of application modules at 
code level, we can obtain more detailed global complexity
$M'_{G}$ of the system: $M'_{G}=  M'_{T} + \sum_{i=1}^{k}M'_{i}.$

In maintenance phases, when we have to modify some component in a
software architecture, usually, we intend to know information about 
how the modified component intersect 
with other components. This kind of information is very useful because
it can tell us if the modified component is a special point that
connects with its environment more closely than other components. 
If so, that means it is difficult to make changes to the component 
due to a large number of potential effects on other components. 
We call such a component the ``most easily affected component of the 
architecture.'' To capture such attribute, we can define following metrics. 

\begin{itemize}
\item Let $D_{t}$ be the set of all dependences arcs in the ADG of a
software architecture, and ${\sigma}_{[1]=v}(D_{t})$
be the number of ports of components on which a port $v$ of a
component is directly dependent. The complexity $M_{S}$ of the most
easily affected component in the architecture can be measured by 
$M_{S}=max \{|{\sigma}_{[1]=v}({D}_{t})|\ | $ {\it v is a vertex of the ADG} \}. 
\end{itemize}

Similarly, if we also considered indirect architectural dependences 
in a software architecture, we can obtain a more detailed 
metric: $M'_{S}=max \{|{\sigma}_{[1]=v}({D}_{t}^+)|\ | $ {\it v is a vertex of the ADG} \}. 

As we observed, all the architectural metrics defined above are
absolute metrics. In general, the larger is a architectural metric of 
a software architecture, the more complex is the software
architecture. Moreover, some relative architectural metrics should
also by considered since they can assess the complexity of a software
architecture from some different viewpoints. 

\section{Related Work}
\label{sec:work}

Although much work has been studied for code metrics at implementation
code level, the study of architectural metrics has not received as
much as attention in comparing with code metrics.
Among existing architectural metrics, there are two famous 
architectural metrics that have been proposed by Yin and Winchester which is
derived from a system's structured design chart, and by Henry and
Kafura which is derived from a system's information flow. We compare 
their approaches with ours here.

Yin and Winchester have defined some architectural metrics based on analysis
of a system's design structure chart \cite{Yin78}. They focused on the
interface between the major levels in a large, hierarchically
structure. However, the fundamental problem for Yin and Winchester's
work is that their metrics were defined based on informal system's
design structure charts which can only capture the flow of information
across level boundaries. In contrast, our metrics are defined based
on various types of architectural dependences in a software
architecture that can be represented by formal 
architectural specification of a system, and therefore, can measure the
architectural complexity of the system more well.

Henry and Kafura proposed some architectural metrics based on information
flow of a system. Their metrics are probably the most cited
architectural metrics that have been developed. The idea behind
these metrics is that complexity is measured in terms of information
flow, and that more complex modules in a system are those through
which large amounts of information flow. Their approach is much more
detailed compared with Yin and Winchester's work because it observes
all information flow rather than just flow across level boundaries.
However, there are two fundamental problems in
information flow metrics. First, although Henry and Kafura stated
that their approach can be completely automated, this is not often the
case. Recent evaluations showed that due to the ambiguous definitions of
some of the metrics, it is difficult to give an evaluation of the
metrics. This makes it difficulty to develop automated support tools
for the approach \cite{Kitchenham90,Shepperd90}. Second, information flow
metrics were also defined based on some informal structure charts
which usually poorly capture the attributes of a system's architecture.
In contrast, our metrics which are defined in terms of various types
of architectural dependences in a software architecture 
that can be represented by formal architectural specification of 
a system, and therefore can capture more intrinsic and deeper
attributes of a system's architecture. Moreover, due to the synthetic
nature of some information flow metrics (i.e. the fact that they are
obtained by combining the values of a number of other counts), recent studies
showed that this may conceal underlying effects and lead to incorrect
diagnoses of the status of either the system as a whole or of
individual components \cite{Kitchenham90}. Our metrics, in contrast,
are defined based on primitive counts (of dependence arcs in the ADG),
rather than synthetics, and therefore no such a problem occurred.

\section{Concluding Remarks}
\label{sec:final}

We proposed some new architectural metrics which are appropriate
for evaluating the architectural attributes of a software system. The main 
feature of our approach is to measure the complexity of a software 
architecture by analyzing various types of architectural dependences 
in the architecture. In order to formally 
define these metrics, we presented a dependence-based representation 
named {\it Architectural Dependence Graph} (ADG) to explicitly
represent these architectural dependences in the architecture. 

The work presented here is primary, and there is still a lot of work 
that remains to be done. For example, in addition to defining metrics based
on architectural dependences, similar to \cite{Bieman94} which they
defined some metrics based on program slices to evaluate functional 
cohesion of a program, we can also define metrics to evaluate the
functional cohesion of a software architecture based on architectural 
slices that can be computed by a new slicing technique called {\it
architectural slicing} \cite{Stafford98,Zhao97,Zhao98}. Moreover, 
we can also define some architectural metrics by simply counting the 
number of elements 
in a formal architectural specification. For example, we can define
metrics by counting the number of components, connections between
components, and even the number of lines in a formal architectural 
specification. 
On the other hand, it is important to develop static analysis tools to
automatically support the collection and evaluation of the
architectural metrics proposed in this paper. Now we are implementing 
an architectural dependence analysis tool for W{\sc right} architectural 
specifications and an architectural metric collector based on it. The
next step for us is to perform some experiments and collect data for 
evaluation. We hope a primary evaluation of these metrics will be 
available soon.

{\small
\renewcommand{\baselinestretch}{0.7}

}

\end{document}